\documentclass[journal,twocolumn]{IEEEtran}
\usepackage{cite}
\usepackage{amsmath,amsfonts,amsxtra,amssymb,latexsym,amscd,amsthm,mathrsfs,bm}
\usepackage{mathtools}
\usepackage{tabularx}
\usepackage{graphicx}
\usepackage{hyperref}
\usepackage{xcolor}
\usepackage{algorithm}
\usepackage{algorithmic}
\usepackage{cite}
\usepackage[colorinlistoftodos,textsize=tiny]{todonotes}
\usepackage{balance}
\usepackage{epstopdf}
\usepackage{balance}
\usepackage{booktabs}
\usepackage{lipsum}

\usepackage{caption}
\usepackage{subcaption}
\newcommand{\bs}{\boldsymbol}
\newcommand{\nn}{\nonumber}

\usepackage{eqparbox}

\usepackage{etoolbox}  
\makeatletter
\patchcmd{\algorithmic}{\addtolength{\ALC@tlm}{\leftmargin} }{\addtolength{\ALC@tlm}{\leftmargin}}{}{}
\makeatother

\newcommand{\T}{\mathsf{T}}

\title{\Large Noncoherent Maximum Likelihood  Detection for LoRa Signals in Multipath Fading 
	\thanks{T. K. Nguyen and E. Bedeer are with the Department of Electrical and Computer Engineering, University of Saskatchewan, Saskatoon, Canada S7N5A9. Emails: \{khai.nguyen,  e.bedeer\}@usask.ca.}
	\thanks{R. Barton is with Cisco Systems Inc.. Email: robbarto@cisco.com.
	}
	\thanks{	This work was supported by the NSERC/Cisco Industrial Research
		Chair.}
}

\author{\IEEEauthorblockN{The Khai Nguyen, Ebrahim Bedeer, and Robert Barton}\vspace{-.5cm}}

\begin{document}
	\maketitle
	\begin{abstract}
		This letter derives the noncoherent (NC) maximum likelihood (ML) detection rule  for LoRa signals under Rician multipath fading channels. The proposed NC-ML detection only requires the channel statistics, not the actual instantaneous channel state information (CSI), which eliminates the overhead associated with channel estimation. Simulation results show that despite the low-complexity, the proposed detection scheme significantly improves the performance of LoRa detection over multipath channels. Notably,  in time-invariant channels, the NC-ML receiver can achieve  equivalent performance as compared to existing coherent schemes, and even surpasses them when Doppler shift is present, while not relying on the channel estimation nor reference signals extracted from the  preamble. 
	\end{abstract}
	\vspace*{-0.45cm}
	\vspace*{-0.5cm}
	\section{Introduction}\label{Sec:Intro}
	Chirp spread spectrum (CSS) modulation, or long range (LoRa) modulation, is one of the key communication standards for Internet of Things (IoT). The constant envelope and linearly-varying-frequency waveform of LoRa signal make it robust against non-linear distortion from power amplifiers and Doppler shift. More importantly, LoRa is also well-known for its power efficiency, which can facilitate long-range-low-power communication \cite{Milarokostas2023} that is  suitable for many IoT applications.
	
	Despite being originally designed for long range, stationary IoT  applications, recently, LoRa has also found its applications in more unfavorable propagation environments with multipath fading channels \cite{Emanuel2019}. In \cite{Demeslay2022,Liu2023}, the theoretical performance of LoRa modulation over multi-path fading channels is derived, which shows that conventional LoRa demodulation scheme has  very poor performance in multipath fading channels. This highlights the importance of designing proper detection schemes that are specialized for multipath fading.
	
	In \cite{Huang2010},  frequency domain equalization schemes for binary orthogonal keying CSS are studied. The paper considers zero forcing (ZF) and minimum mean squared error (MMSE) equalizers, in which the MMSE is shown to outperform the ZF at the cost of  higher complexity.
	In \cite{Guo2020}, a time-delay-estimation-like (TDEL) detector for LoRa in multi-path fading is proposed. Based on the fact that any received LoRa chirp  is  a circularly shifted version of one another, the authors  perform cyclic correlation between the received signal and the preamble to detect the transmitted chirp. 
	In \cite{Demeslay2022b}, a matched filter approach was proposed, where  multipath interference can be combined constructively similar to a RAKE receiver structure. The reported results show superior  performance as compared to previous coherent  designs for multipath channels. Unlike the aforementioned works, which focus on  \emph{receiver} design for multipath channels, in \cite{Liu2024}, a block interleaved chirp spreading LoRa  \emph{transmission} scheme is proposed to reduce the probability that the interfering peaks from non-direct paths surpass the peak corresponding to the direct path. It is shown that the performance gain of this transmission scheme grows as the spreading factor (SF) increases.
	
	The aforementioned works  require channel state information (CSI), which is  estimated from LoRa  preamble \cite{Demeslay2022b}, except for \cite{Guo2020}, which does not explicitly require channel estimation, but still needs the CSI in the form of a reference signal extracted from the preamble. Not only is this a complexity overhead, but it also makes the detector susceptible to errors caused by the mismatch between the estimated and the true instantaneous CSI. The reason is that a LoRa  packet has a maximum payload from $51$ (SF12) to $222$ bytes (SF7) \cite{Milarokostas2023}, and  thus,  the channel is unlikely to stay constant over such  significantly long airtime  in the presence of Doppler effect. For example, a LoRa packet with 500-kHz bandwidth, 222-byte payload and 13-byte header  has an airtime of $92.2$, $164.9$, $292.1$, $533.0$ ms for $\mathrm{SF}=7$, 8, 9, 10, respectively \cite{semtech}.  With end device (ED)   mobility of $\nu = 1$ m/s (3.6 km/h) at $f_c=915$ MHz carrier frequency, the Doppler spread is $f_d = \nu/(3\times10^8)\times f_c \approx 3$ Hz, and the coherence time is $\tau_c\approx 1/(4f_d) = 81$ ms, which is well below LoRa packets' airtime. Thus,  the channel will vary significantly over the packet duration, and hence, the  CSI based on the preamble can hardly remain valid over the  span from the preamble to the entire payload.
	
	Therefore, in this paper, we derive a low-complexity noncoherent (NC) maximum likelihood (ML) detection rule for LoRa signals in multipath Rician fading channels, which is the general and practical scenario in LoRaWAN, and can characterize both line-of-sight (LoS) and non-LoS transmission by simply adjusting the shape factor of the Rician distribution. Despite the low-complexity, it will be shown with simulation results that the proposed NC-ML detection has  equivalent performance as the preamble-based TDEL method in \cite{Guo2020} in time-invariant channel, and can be even better  in the presence of Doppler effect,  while not being reliant on \emph{instantaneous} CSI acquisition like existing detection candidates.
	
	The remainder of this letter is organized as follows. Section~\ref{sec:Sys} presents the system model with multipath fading. Section~\ref{sec:Detection} derives the coherent and NC-ML LoRa detection rules for Rician multipath fading channel. Section~\ref{sec:sim} provides simulation results. Finally, Section~\ref{sec:con} concludes the paper.
	
	\textit{Notations:} The convolution  is denoted by $\odot$. The circular convolution is denoted by $\circledast$. Transposed matrices are denoted by the superscript $\T$. Conjugate transpose of a complex number is denoted by the superscript $*$. Identity matrix of size $N$ is denoted by $\mathbf{I}_N$. The modified Bessel function of the first order is $I_0(\cdot)$. The modulo operator is denoted by $\mathrm{mod}$. The set of $d_1\times d_2$ complex matrices is  $\mathbb{C}^{d_1\times d_2}$. The complex normal distribution with variance $\sigma^2$ is  $\mathcal{CN}\left(0,\sigma^2\right)$. The Rayleigh distribution with the scale parameter $\sigma$ is  $\text{Rayleigh}(\sigma)$. The Rician distribution with distance parameter $\nu$ and scale parameter $\sigma$ is  $\text{Rice}(\nu,\sigma)$. The real and imaginary parts of a complex number are  $\mathfrak{R}\{\cdot\}$ and $\mathfrak{I}\{\cdot\}$, respectively.
	\section{System Model}\label{sec:Sys}
	A LoRa packet is constructed from multiple symbols, which are drawn from a  set of $M=2^{\mathrm{SF}}$ LoRa   chirps.  Specifically, the $m$th chirp is defined as $\bs{s}_{{m}}=[s_{{m}}[0],s_{{m}}[1],\ldots,s_{{m}}[M-1]]^{\T}$, where ${m}=0,1,\ldots,M-1$. The $0$th chirp is known as the basic upchirp, whose samples are defined as
	$
	s_{0}[n]=\exp\left\{j2\pi\left(\frac{ n^2}{2M}-\frac{n}{2}\right)\right\},\;n=0,1\dots M-1.
	$
	Then, the $n$th sample of the $m$th chirp can be simply obtained from $s_m[n]$ as $s_m[n]=s_0[(n+m)\mathrm{mod}M]$.	
	
	Consider the transmission of one \emph{single} LoRa symbol $\boldsymbol{s}_m$ over a multipath channel of $L$ taps $\boldsymbol{h}=[h_0,h_1\dots h_{L-1}]^\T$. To make the channel model practical for LoRa, the magnitude of the first tap is assumed to follow a Rician distribution with the shape factor $K_0$, since it may contain the LoS path (when the LoS path is blocked, $K_0=0$). Whereas, the magnitudes of other taps are assumed to follow Rayleigh distributions (no LoS path, equivalent to a Rician distribution with the shape factor $K_{\ell}=0,\;\ell\neq0$). In general,  $|h_{\ell}|\sim\text{Rice}\left(\sqrt{K_{\ell}\rho_{\ell}/(K_{\ell}+1)},\sqrt{\rho_{\ell}/(K_{\ell}+1)}\right)$, 	where $\rho_{\ell}$ is the average  power of the $\ell$th tap, with the probability distribution function (pdf)   \cite{goldsmith}:
	\begin{align}
		f(|h_{\ell}|)&=\frac{2(K_{\ell}+1)|h_{\ell}|}{\rho_{\ell}}\exp\left\{-K_{\ell}-\frac{(K_{\ell}+1)|h_{\ell}|^2}{\rho_{\ell}}\right\}\nn\\
		&\times I_0\left\{2\sqrt{\frac{K_{\ell}(K_{\ell}+1)}{\rho_{\ell}}}|h_{\ell}|\right\}.
	\end{align}
	
	The received signal over the channel is then given by:
	\begin{equation}
		\boldsymbol{y} = \boldsymbol{h}\odot\boldsymbol{s}_m + \boldsymbol{w} \in\mathbb{C}^{(M+L-1)\times 1},
	\end{equation}
	where  $\boldsymbol{w}\sim\mathcal{CN}(0,\sigma^2\mathbf{I}_{M+L-1})$ is AWGN noise. 
	In practice, due to the narrow-bandwidth (125 to 500 kHz),  LoRaWAN channel usually has very few taps, which is calculated as $L= \lceil\tau_{\mathrm{max}}\times B\rceil$, where $\tau_{\mathrm{max}}$ is the maximum delay.  For example, with the extended vehicular A channel model  \cite{ITU} with $\tau_{\mathrm{max}}=2.5$  $\mu s$ and  500-kHz bandwidth, the channel has $L=2$ taps. This number  is negligible compared to a LoRa symbol length ($M=128$ to $4096$ samples for SF7 to SF12). Since $L<<M$, the effect of inter-symbol interference (ISI) is negligible and can be ignored, which is a common assumption in the literature \cite{Demeslay2022b,Guo2020}. As a result, the input-output signal relationship over multipath fading channels can be approximated  as:
	\begin{align}
		\tilde{\boldsymbol{y}} &= \boldsymbol{h}\circledast\boldsymbol{s}_m + \tilde{\boldsymbol{w}} \in\mathbb{C}^{M\times 1}
	\end{align}
	where  $\tilde{\boldsymbol{w}}\sim\mathcal{CN}(0,\sigma^2\mathbf{I}_{M})$. The elements of $\tilde{\boldsymbol{y}}$ can be rewritten:
	\begin{align}
		\tilde{{y}}[n] &= \sum_{\ell=0}^{L-1}h_{\ell}s_{m}[(n-\ell)\mathrm{mod}M] +\tilde{w}[n]\nn\\
		&=\sum_{\ell=1}^{L}h_{\ell}s_{(m-\ell)\mathrm{mod}M}[n]+\tilde{w}[n].
	\end{align}
	This shows that  when a single LoRa symbol is transmitted over multipath  channel, it appears as the sum of $L$ LoRa symbols, where each of those corresponds to one tap of the channel. Consequently, if the conventional detector is used \cite{Demeslay2022,Liu2023}, which implements dechirping, followed by the discrete Fourier transform (DFT), the received dechirped signal in the frequency domain is $\tilde{\boldsymbol{Y}}=[\tilde{Y}[0],\tilde{Y}[1]\dots\tilde{Y}[M-1]]^{\T}$, where:
	\begin{equation}\label{fft}
		\begin{split}
			\tilde{Y}[k]
			=\begin{cases}
				\sqrt{M}{h}_{\ell}e^{j\psi_{k}} + \tilde{W}[k],\;&k=(m-\ell)\mathrm{mod}M\\
				\tilde{W}[k],&\text{otherwise}.
			\end{cases} 
		\end{split}
	\end{equation}\linebreak
	In \eqref{fft}, $\tilde{W}[k]$ is the DFT of the noise $\tilde{\boldsymbol{w}}$ measured at the $k$th frequency bin and
	$
	\psi_{{k}}=2\pi\left\{{ {{k}}^2}/(2M)-{{{k}}}/{2}\right\}$ is a deterministic bin-dependent phase offset, which can be conveniently discarded by converting $\tilde{Y}[k]$ into:
	\begin{equation}
		\bar{Y}[k]=\tilde{Y}[k]\exp\{-j\psi_{{k}}\}.
	\end{equation}
	Equation \eqref{fft} indicates that over multipath channel, the dechirped signal in the frequency domain allocates the energy on $L$ different bins. Consequently, since the conventional DFT-based receiver detects symbols by selecting the bin with the highest power, it is extremely susceptible to errors due to multiple unwanted peaks at different  bins following the one corresponding to the transmitted symbol. Thus, in the next section, we propose a low-complexity  NC-ML detection method for LoRa signal in multipath channels.

			\section{LoRa ML detection for multipath fading}\label{sec:Detection}
			In this section, the NC-ML detection rule for LoRa signal over Rician multipath fading is derived. The coherent ML detection rule will also be derived for comparison purposes.
			\subsection{Coherent maximum likelihood detection}
			According to \eqref{fft}, given that the $m$th symbol has been transmitted, it can be seen that conditioned on $h_{\ell}$:
			\begin{align}\label{eq.cohDis}
				&\bar{Y}[k]\sim\mathcal{CN}(\sqrt{M}h_{\ell},\sigma^2),\; &k=(m-\ell)\mathrm{mod}M,\nn\\
				&\bar{Y}[k]\sim\mathcal{CN}(0,\sigma^2),&\text{otherwise}.
			\end{align}
			Consequently, the likelihood function that the $m$th chirp is transmitted, given the received signal $\bar{\boldsymbol{Y}}$, is the product of the pdf of all $M$ frequency bins with the distributions in \eqref{eq.cohDis} conditioning on the channel $\boldsymbol{h}$:
			{\small
				\begin{align}\label{eq.LLH-C}
					&\mathcal{L}(m|\bar{\boldsymbol{Y}},\boldsymbol{h})=\left(\frac{1}{\pi\sigma^2}\right)^M\left(\prod_{\substack{k = 0\\ k \neq (m-\ell)\mathrm{mod}M}}^{M-1}\exp\left\{-\frac{|\bar{Y}[k]|^2}{\sigma^2}\right\}\right)\nn\\
					&\times\left(\prod_{L-1}^{\ell=0}\exp\left\{-\frac{|\bar{Y}[(m-\ell)\mathrm{mod}M]-\sqrt{M}h_\ell|^2}{\sigma^2}\right\}\right)\nn\\
					&=\frac{\prod_{k=0 }^{M-1}\exp\left\{\frac{-|\bar{Y}[k]|^2}{\sigma^2}\right\}}{(\pi{\sigma^2})^M}\prod_{L-1}^{\ell=0}\frac{\exp\left\{\frac{-|\bar{Y}[(m-\ell)\mathrm{mod}M]-\sqrt{M}h_\ell|^2}{\sigma^2}\right\}}{\exp\left\{-\frac{|\bar{Y}[(m-\ell)\mathrm{mod}M]|^2}{\sigma^2}\right\}}.\nn\\
			\end{align}}
			Taking the logarithm of the likelihood function results in the log-likelihood (LLH) function in a simpler form as:
			{\small
				\begin{align}
					&\mathrm{LLH}(m|\bar{\boldsymbol{Y}},\boldsymbol{h})=\mathrm{log}(\mathcal{L}(m|\bar{\boldsymbol{Y}}),\boldsymbol{h})\nn\\
					&=C+\frac{\sum_{\ell=0}^{L-1}\left(\sqrt{M}\mathfrak{R}\{\bar{Y}[(m-\ell)\mathrm{mod}M]h_{\ell}^{*}\} +M|h_{\ell}|^2\right)}{\sigma^2},
				\end{align}
			} \linebreak where $
			C=\mathrm{log}\left(\left(\frac{1}{\pi{\sigma^2}}\right)^M\prod_{k=0 }^{M-1}\exp\left\{-\frac{|\bar{Y}[k]|^2}{2\sigma^2}\right\}\right)$ and is independent of $m$. From this LLH function, the coherent ML detection rule can be expressed as:
			\begin{align}\label{eq.COHrule}
				\hat{m}=\underset{m}{\mathrm{argmax}}\;\sum_{\ell=0}^{L-1}\mathfrak{R}\{\bar{Y}[(m-\ell)\mathrm{mod}M]h_{\ell}^{*}\}.
			\end{align}
			It can be seen   that the obtained coherent ML detection rule  above is identical to the RAKE receiver structure  in \cite{Demeslay2022b}.

			\subsection{Noncoherent  maximum likelihood detection}
			
			The likelihood function that the $m$th chirp is transmitted  with the NC-ML detection can be obtained by marginalizing the coherent likelihood function in \eqref{eq.LLH-C} over the channel $h_{\ell}$.  However, this approach involves the calculation of  $L$-fold challenging integrals that would make the derivation lengthy and hard to follow. Note that  marginalizing  the conditional bin distribution in \eqref{eq.cohDis}  over $h_{\ell}$  results in the bin distribution in the NC case (not conditioned on $h_{\ell}$). Consequently, the NC likelihood function obtained by marginalizing the coherent likelihood function \eqref{eq.LLH-C} over $h_{\ell}$ will be equal to the product of the pdfs of the received bins' magnitudes. Thus, we chose to  derive the NC-ML directly from the distributions of the magnitudes of the frequency bins, which is more straightforward.

			With Rician multipath fading, the LoS path, which contributes to  the first channel tap, has deterministic gain, while other paths (non-LoS) are assumed to have magnitude following Rayleigh distribution. 
			Given the $m$th chirp was transmitted, the magnitude of the received signal in the frequency domain in \eqref{fft} after DFT will be distributed as:
			\begin{align}\label{eq.NCdis}
				&|\bar{Y}[k]|\sim\text{Rice}\left(\sqrt{\frac{K_0M\rho_0}{K_0+1}},\sqrt{\frac{M\rho_0}{K_0+1}+\sigma^2}\right),\;k=m, \nn\\
				&|\bar{Y}[k]|\sim\text{Rayleigh}(\sqrt{M\rho_{\ell}+\sigma^2}),\; k=(m-\ell)\mathrm{mod}M,\nn\\
				&|\bar{Y}[k]|\sim\text{Rayleigh}(\sigma),\;\text{otherwise}.
			\end{align}
				\textit{Proof:} When $k\neq m$, from \eqref{fft}, it can be  seen that $\bar{Y}[k]\sim\mathcal{CN}(0,M\rho_{\ell}+\sigma^2)$ for $k=(m-\ell)\mathrm{mod}M$ and $\bar{Y}[k]\sim\mathcal{CN}(0,\sigma^2)$ otherwise. As a result, the amplitude of $\bar{Y}[k]$ follows Rayleigh distribution as in \eqref{eq.NCdis}.
				When $k=m$,
				$
				\bar{Y}[k]=\sqrt{M}{h}_{\ell} + \bar{W}[k]
				$.
				Since ${h}_{\ell}$ is Rician distributed with the shape factor $K_0$ and raw moment $\rho_{0}$, the first term  is also a Rician distributed  with the shape factor $K_0$ and raw moment $M\rho_{0}$. Thus, when adding the AWGN noise $\bar{W}[k]\sim\mathcal{CN}(0,\sigma^2)$,  $\bar{Y}[k]$ will also follow a Rician distribution, but with the new raw moment $M\rho_{0}+\sigma^2$. As a result, the new shape factor $\tilde{K}_{0}$ of the first tap can be easily calculated as $
				\tilde{K}_0 = {K_0M\rho_0}/(M\rho_0+(K_0+1)\sigma^2)
				$. $\hfill\blacksquare$

			Consequently, the NC-ML detection can be derived as:
			\begin{align}\label{eq.ruleNC}
				&\hat{m}=\underset{m}{\mathrm{argmax}}\;\left(\mathrm{log}\left\{I_0\left\{2\sqrt{\frac{\tilde{K}_0(\tilde{K}_0+1)}{M\rho_{0}+\sigma^2}}|\bar{Y}[m]|\right\}\right\}\sigma^2\right.\nn\\
				&\left.+\frac{M\rho_{0}|\bar{Y}[m]|^2}{M\rho_{0}+({K}_0+1)\sigma^2}+\sum_{\ell=1}^{L-1}\frac{M\rho_{\ell}|\bar{Y}[(m-\ell)\mathrm{mod}M]|^2}{M\rho_{\ell}+\sigma^2}\right).
			\end{align}

			\textit{Proof:} With the distribution in \eqref{eq.NCdis}, the likelihood function can be calculated  as the product of the pdf of all $M$  bins, which is presented in  \eqref{eq.LLNC} on top of the next page. By eliminating the terms that are independent of $m$ and taking the logarithm, the NC-ML rule can be easily obtained as in  \eqref{eq.ruleNC}. $\hfill\blacksquare$

			As mentioned earlier, the NC-ML detector requires the knowledge of the channel's parameters, which can be estimated \emph{offline} based on the preamble of multiple packets received over time with many well-investigated parameter estimation techniques such as the low-complexity moment-based method in \cite{Bakkali2007}, where the estimated parameters $K_{\ell}$ and $\rho_{\ell}$ can be found via solving a quadratic equation with coefficients as the second-order and forth-order average values of the observation data (Eq.~(13) of \cite{Bakkali2007}).  Since the EDs for LoRa applications are mostly stationary or of very low-mobility, the channel parameters $K_{\ell}$ and $\rho_{\ell}$ vary slowly and slightly. Thus, there is no need to perform parameter estimation after every received packet. Instead, after every received packet, the channel information can be extracted from the preamble and stored in a data set for future parameter estimation (after a predetermined number of received packets, or a period of time, depending on the applications and/or environments).
			
			\begin{figure*}[t!]
				\small
				\vspace{-.6cm}
				\begin{align}\label{eq.LLNC}
					\mathcal{L}(m|\bar{\boldsymbol{Y}})&=\left(\prod_{\ell=1}^{L-1}\frac{\exp\left\{-\frac{|\bar{Y}[(m-\ell)\mathrm{mod}M]|^2}{M\rho_{\ell}+\sigma^2}\right\}}{\exp\left\{-\frac{|\bar{Y}[(m-\ell)\mathrm{mod}M]|^2}{\sigma^2}\right\}}\right)\frac{\exp\left\{-\frac{(\tilde{K}_0+1)|\bar{Y}[m]|^2}{M\rho_{0}+\sigma^2}\right\}}{\exp\left\{-\frac{|\bar{Y}[m]|^2}{\sigma^2}\right\}}I_0\left\{2\sqrt{\frac{\tilde{K}_0(\tilde{K}_0+1)}{M\rho_{0}+\sigma^2}}|\bar{Y}[m]|\right\}\nn\\
					&\times\underbrace{\left(\prod_{k=0}^{M-1}\frac{2|\bar{Y}[k]|}{\sigma^2}\exp\left\{-\frac{|\bar{Y}[k]|^2}{\sigma^2}\right\}\right)\left(\prod_{\ell=1}^{L-1}\frac{\sigma^2}{M\rho_{\ell}+\sigma^2}\right)\exp\left\{-\tilde{K}_0\right\}\frac{(\tilde{K}_0+1)\sigma^2}{M\rho_{0}+\sigma^2}}_{\text{independent of $m$}}.
				\end{align}
				\rule{\linewidth}{.1pt}
				\vspace{-0.9cm}
			\end{figure*}

			\subsection{Discussion}

				Unlike the conventional LoRa receiver, which does not account for the multipath channels that spread  power across multiple bins, the NC-ML receiver, the coherent ML receiver, and the TDEL method combine the power dispersed over these bins so that the bin corresponding to the transmitted  symbol attains the highest power. The difference resides in the manner the bins are weighted before combination. The coherent ML weights the bins with instantaneous channels, which yields the highest signal to noise ratio (SNR) \cite{Demeslay2022b}, while the TDEL uses preamble power as weight coefficients. Whereas, the NC-ML weights the bins with the statistical average power. Thus, it appears that the coherent ML and the TDEL receiver, with the instantaneous CSI, would be the better options for time-invariant channel, compared to the NC-ML detector, which relies on the statistics of the channels, and might fail with tail-event channel behavior (for example, the first tap channel is in deep fade while the delay taps are strong). However, in time-variant channels, the NC-ML can be  better, since the CSI for the coherent ML and the TDEL becomes outdated  when the packet airtime is beyond channels' coherence time.
				
				\subsection{Complexity analysis}
				The steps of the proposed NC-ML detection and its complexity are summarized as follows:
				\begin{enumerate}
					\item Dechirp and DFT: complexity of $\mathcal{O}(M\mathrm{log}_2M)$.
					\item Calculate the decision metric in \eqref{eq.ruleNC}: $M$  functions of $M$  chirps are evaluated. For each one, a sum of $L$ elements must be performed, which results in the complexity order of $\mathcal{O}(ML)$ \footnote{Note that the  function $I_0(\cdot)$ has constant time complexity when evaluated via a fixed-length power series or look-up table.}
					\item Find the bin with the highest likelihood function among $M$ bins requires $\mathcal{O}(M)$ complexity.
				\end{enumerate} 
				Together, the total complexity of the coherent and NC-ML detection is $\mathcal{O}(M\mathrm{log}_2M+ML)$, which is equivalent to the complexity of the TDEL method \cite{Guo2020} and the coherent ML receiver (or RAKE receiver \cite{Demeslay2022b}).
			\section{Simulation Results}\label{sec:sim}
\begin{figure}[t!]
				\centering
				\includegraphics[width=0.5\textwidth]{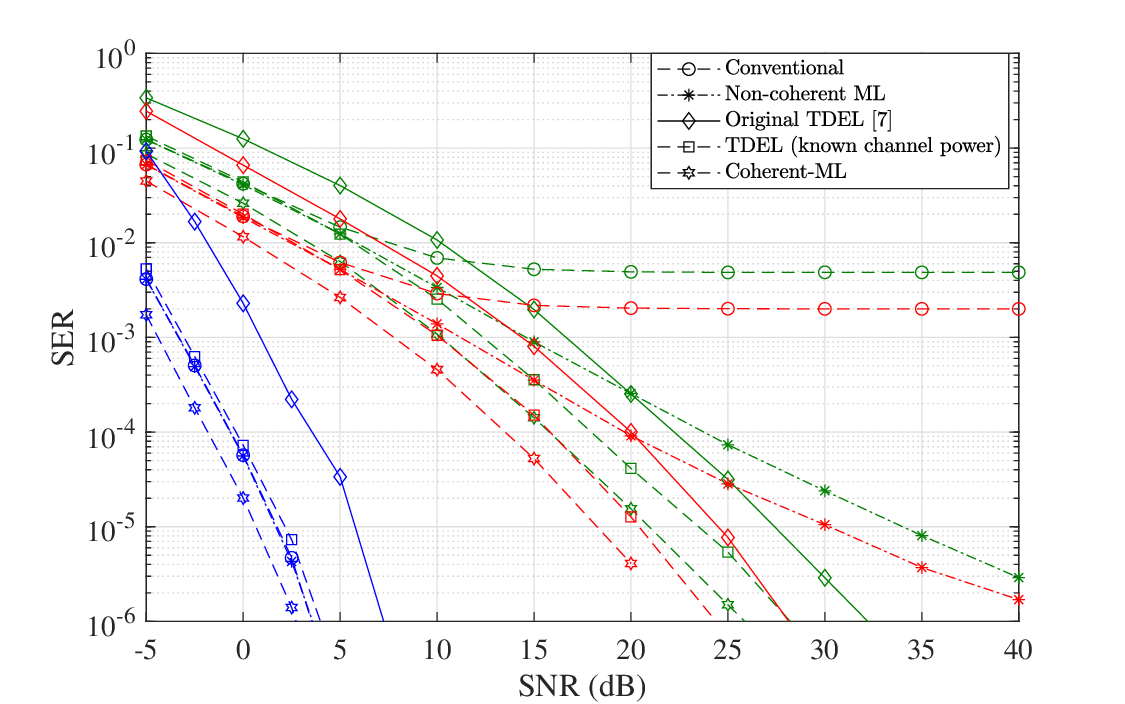}
				\caption{SER in time-invariant channel ($f_d=0$ Hz). Green, red, blue are for $K_0=0, 2, 10$, respectively ($\mathrm{SF}=7$).}
				\label{fig_ser_noDoppler}
			\end{figure}

\begin{figure}[t!]
				\centering
				\includegraphics[width=0.5\textwidth]{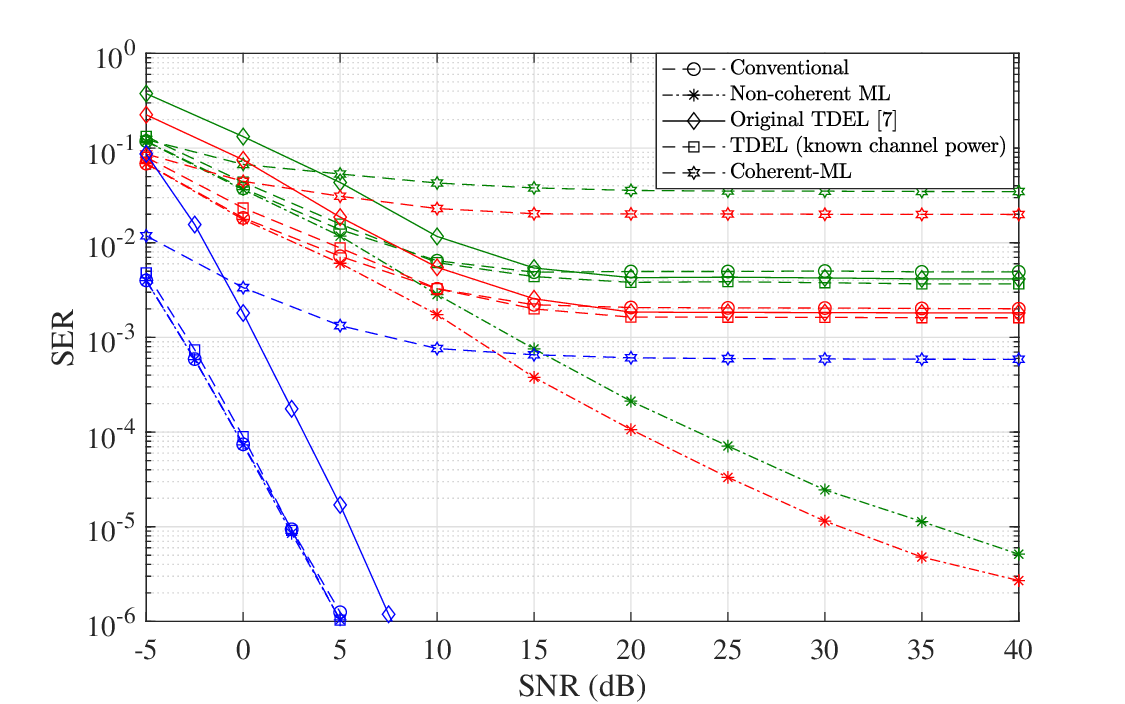}
				\caption{SER in time-variant channel ($f_d=5$ Hz). Green, red, blue are for $K_0=0, 2, 10$, respectively ($\mathrm{SF}=7$).}
				\label{fig_ser_Doppler5Hz}
			\end{figure}

\begin{figure}[t!]
				\vspace{-.3cm}
				\centering
				\includegraphics[width=0.5\textwidth]{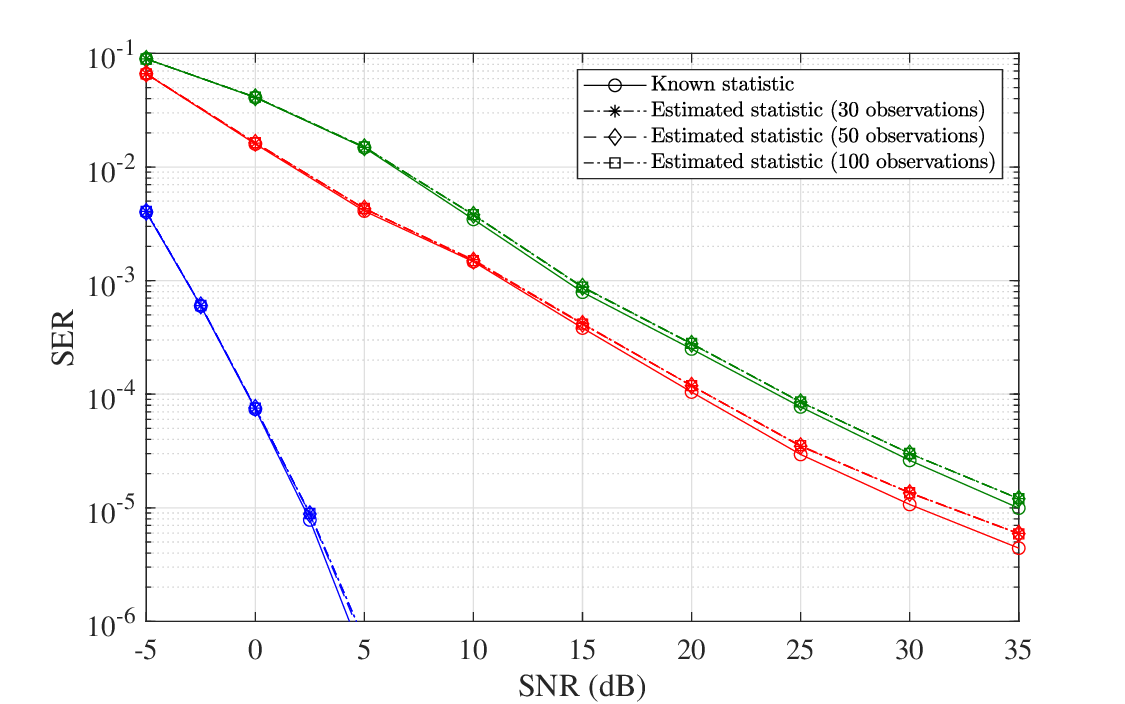}
				\caption{SER with perfect and estimated parameters. Green, red, blue are for $K_0=0, 2, 10$, respectively ($f_d=5$ Hz$, \mathrm{SF}=7$).}
				\label{fig_param_estimation}
			\end{figure}

			This section presents the performance comparison among the NC-ML detection, the conventional LoRa detection, the TDEL detection method \cite{Guo2020}, which has the best NC detection performance in recent literature, and the coherent ML detection (which is identical to the  receiver  in \cite{Demeslay2022b}). Note that for the original TDEL method  \cite{Guo2020}, the reference signal is extracted from the preamble by setting the amplitudes of the frequency bins that are lower than $1/4$ of the highest bin's amplitude to zero. This makes the original TDEL method somewhat susceptible to noise. Thus, for a fair comparison, we also present the performance of a modified TDEL method, which knows the number of channel taps in advance, and thus, can set the  bins of the reference signal that do not correspond to the channel power profile to zero to reduce noise.  The CSI is assumed to be perfectly known  for the coherent ML detection \cite{Demeslay2022b}. The results for the NC-ML detection are obtained with  known channel parameters in Figs.~\ref{fig_ser_noDoppler}, and \ref{fig_ser_Doppler5Hz}, and estimated channel parameters ($K_{\ell}$, $\rho_{\ell}$) in Fig.~\ref{fig_param_estimation}.
			
			The extended vehicular A model \cite{ITU} with the channel delay profile $\boldsymbol{\tau}=[0,\, 30,\, 150,\, 310,\, 370,\, 710,\, 1090,\, 1730,\, 2510]$ nanoseconds and the respective relative power profile of $[0,\, -1.5,\, -1.4,\, -3.6,\, -0.6,\, -9.1,\, -7,\, -12,\, -16.9]$ dB is used. To simulate the most unfavorable multipath scenario, we chose SF7 for the shortest symbol period, and  the bandwidth $B=500$ kHz for the largest number of taps. Consequently, the channel filter has $L=\lceil\mathrm{max}(\boldsymbol{\tau})\times B\rceil=2$ taps. The carrier frequency is 915 MHz. The packet payload length is  $100$ LoRa symbols.
			The results will be obtained with  Rician fading shape factors of the first channel tap $K_0=0$ (Rayleigh fading), $2$ and $10$. The rate of channel variation  is determined by  the maximum Doppler spread $f_d$.
			
			In Fig.~\ref{fig_ser_noDoppler}, the symbol error rates (SER) of the  aforementioned detection methods  in time-invariant fading channel ($f_d=0$ Hz) are shown. The SER of the conventional LoRa detection quickly reaches an error floor when $K_0=0$ and $2$. Whereas, with the NC-ML  and the TDEL methods, the SER continues to decrease as the SNR increases. Without the knowledge of the channel power profile, the original TDEL method is worse than the NC-ML detection in the low SNR range, then surpasses it around the SNR of $21$ dB. Meanwhile, with the knowledge of the channel power profile, the TDEL method can achieve a power gain of 5 dB as compared to the NC-ML detection at the SER of $10^{-4}$ when $K_0=0$ or $2$, while still being worse than the NC-ML detection at the low SNR range (below $5$ dB). By contrast, when $K_0=10$, which is close to  a channel with constant LoS path gain, the performance of the NC-ML detection is almost identical to the conventional LoRa detection, which is closely followed by the TDEL method (roughly 0.5 dB loss at the SER of $10^{-4}$). Without the knowledge of the channel power profile for the TDEL, this gap  widens to 4 dB. The coherent ML receiver is the best detector in time-invariant channels, with a gain of around 2 dB (at the SER of $10^{-4}$) over other schemes, thanks to the ability to eliminate the noise on the imaginary part of the signal in \eqref{eq.COHrule}.
			
			In Fig.~\ref{fig_ser_Doppler5Hz}, the performance comparison obtained in slowly-time-variant channel  is presented. The maximum Doppler  is set at $f_d=5$ Hz, which can be experienced with a moving LoRa device at a  walking speed of $6$ kms per hour. According to the figure, the TDEL method's performance when $K_0=0$ or $2$, whether the channel power profile is available or not, reaches an error floor at the high SNR range. This is due to the time-variant channel, which makes the reference signal taken from the preamble  outdated over time, especially at the last symbols of  LoRa packets. The only exception is when $K_0=10$, where the deterministic, direct LoS path is dominant, and is only phase shifted by the Doppler effect. In this case, the channel power (especially the first tap) experiences very small variation over time (since the value of $K_0$ is large). Therefore, the reference signal extracted from the preamble is still valid over the long duration of the payload. Whereas, the performance of the conventional LoRa detection and the proposed NC-ML detection is almost identical to the time-invariant case, since the instantaneous CSI is not required. Thus, the NC-ML detection outperforms the TDEL and the conventional LoRa detection in time-variant channel.

			Noticeably, the coherent ML detection has the worst performance among the evaluated methods in time-variant channel, due to the strong reliance on the accuracy of CSI. Its SER quickly reaches an error floor even before the conventional LoRa detection. To maintain  good performance as compared to the time-invariant channel case, the TDEL and the coherent ML methods will need frequent reference updates/channel estimations within a LoRa packet to update the instantaneous CSI. This, however, comes at the cost of a significant training overhead that reduces the already low data rate of LoRa modulation. Thus, in the time-variant channel scenario, NC-ML detection appears to be the most suitable approach to detect  LoRa signals over multipath fading.
			
				Fig.~\ref{fig_param_estimation} compares the error performance of the NC-ML detector with perfect and estimated channel parameters. The parameters  $\rho_{\ell}$ are assumed to   vary uniformly by $10\%$ around their average value from packet to packet, which accounts for the movements of the ED or  transmission environment (moving reflecting objects and obstacles)  relative to the GW, while remaining constant within the packet duration (since the movement within milliseconds of a packet duration is negligible). It can be seen that with  30 observations and above, the performance appears to converge, which  is around 2-3 dB behind the performance with perfectly known parameters when $K_{0}=2$ and $K_{0}=0$. Meanwhile, with $K_{0}=10$, there is almost no difference.  For a LoRa system with SF7, these 30 observations (packets) only takes roughly 3 seconds of airtime,  and can be refined along the way in an \emph{offline} manner when more packets are received. This shows the resilience of the proposed method against slight environment variation.

			\section{Conclusions}\label{sec:con}
			
			This letter derives the NC-ML detection rule for LoRa signals in multipath Rician fading channels. The NC-ML detection does not require the overhead associated with channel estimation, and only requires the knowledge about the statistics of the channel. Despite the low-complexity, the proposed NC-ML detection can significantly improve the SER as compared to the conventional LoRa receiver.  Moreover,
			in the time-invariant channel scenario,  with a weak LoS path ($K_0=0$ to $2$) the NC-ML is 2 dB better than the original TDEL method, while being around 2-5 dB worse than the modified TDEL and the coherent ML detection at the SER of $10^{-4}$. Whereas, with a strong LoS path, the NC-ML can achieve roughly $4$ and $0.5$ dB gain over the original and modified TDEL methods, respectively, while still being around 2 dB behind the coherent ML. By contrast, in the presence of Doppler shift, the  NC-ML detection even outperforms the TDEL methods  and coherent ML detection, which quickly approach error floors (well above the SER of $10^{-4}$) when the LoS path is weak ($K_0=0$ and 2). This is largely   thanks to the independence from the CSI requirement of the NC-ML detection. With a strong LoS path ($K_0=10$), the proposed NC-ML detection  outperforms both the original and modified TDEL by $4$ and $0.5$ dB, respectively, while the coherent ML still reaches an error floor. These results indicate that the proposed NC-ML detection is a suitable receiver for LoRa in multipath channels, especially with the presence of Doppler shift.

			\bibliographystyle{IEEEtran}

		\end{document}